\documentclass[twocolumn,showpacs,preprintnumbers,amsmath,amssymb]{revtex4}
\usepackage{graphicx}
\usepackage{lipsum}
\begin{document}
\title{The role of spin-flip assisted or orbital mixing tunneling on transport through strongly correlated multilevel quantum dot}
\author{D. Krychowski and S. Lipi\'nski}
\affiliation{%
Institute of Molecular Physics, Polish Academy of Sciences\\M. Smoluchowskiego 17,
60-179 Pozna\'{n}, Poland
}%
\date{\today}
\begin{abstract}
Using  the slave boson Kotliar-Ruckenstein approach (SBMFA) for N level Anderson model, we compare fully symmetric SU(N) Kondo resonances occurring for spin and orbital conserving tunneling  with many-body resonances for the  dot with broken symmetry caused by spin, orbital or full spin-orbital mixing. As a result of interorbital  or spin flip processes new interference paths emerge, which manifests in the occurrence of antibonding Dicke like and bonding Kondo like resonances. The analytical expressions for linear conductances and linear temperature thermopower coefficient for arbitrary  N are found.\end{abstract}
\maketitle

\section{Introduction}
The growing  interest in the  fundamental many-body phenomenon - Kondo effect is  stimulated  not only by the  purely cognitive purposes, but  also by a rich of potential applications in quantum electronics. The spin  SU(2)  Kondo effect was first observed on a nanoscopic scale in semiconductor quantum dots (QD) \cite{Kastner}. The SU(N) Kondo physics with $N = 4$ is experimentally realized in carbon nanotubes \cite{Herrero} and double QDs \cite{Goldhaber}.  Suggestions for realizations of SU(N) with $N = 3$ can be found e.g. in \cite{Moca, Krychowski}, $N = 6$ in \cite{Kuzmenko} and $N = 12$ in \cite{Avishai}. In this article we examine the impact of  spin or orbital pseudospin flip processes associated with tunneling on many-body  resonances and present  how this is reflected in transport properties.

\section{Model and formalism}
We consider multiorbital quantum dot or a set of QDs described by generalized $N$-orbital Anderson model:
\begin{eqnarray}
&&{\cal{H}}={\cal{H}}_{d}+{\cal{H}}_{c}+{\cal{H}}^{dir}_{d-c}+{\cal{H}}^{mix}_{d-c}
\end{eqnarray}
where ${\cal{H}}_{d}=\sum_{ls}E_{d}n_{ls}+{\cal{U}}\sum_{l\neq l'ss'}(n_{l\uparrow}n_{l\downarrow}+n_{ls}n_{l's'})$
is hamiltonian of the dot with single- particle energy $E_{d}$ and Coulomb interaction (${\cal{U}}$), ${\cal{H}}_{c}=\sum_{k\alpha ls}E_{k}n_{k\alpha ls}$ describes  electrodes ($\alpha=L,R$). ${\cal{H}}^{dir}_{d-c}$ represents direct tunneling processes ${\cal{H}}^{dir}_{d-c}={\cal{V}}\sum_{k\alpha ls}(c^{\dagger}_{k\alpha ls}d_{ls}+h.c.)$ and the mixing term reads ${\cal{H}}^{mix}_{d-c}={\cal{V}}'\sum_{k\alpha lsl's'}(c^{\dagger}_{k\alpha ls}d_{l's'}+h.c.)$ where $s'\neq s$ or $l'\neq l$. In general the   direct hopping integral ${\cal{V}}$ differs from  mixing hopping integral  ${\cal{V}}'$. The degree of mixing will be characterized by parameter $\nu={\cal{V}}'/{\cal{V}}$, $0\leq\nu\leq1$.

In the present paper we compare transport properties of  the fully symmetric SU(N) systems ($nm$-no mixing between the channels) (${\cal{V}}'=0$) with transport in the following cases explained on Fig. 1a: tunneling conserving only spin - orbital mixing ($o$-different orbital channels are mixed), $(\alpha kls)\leftrightarrow (l's)$ where $l'\neq l$, tunneling conserving only orbital quantum numbers - spin mixing ($s$), $(\alpha kls)\leftrightarrow (l,-s)$ and the case when mixing occurs  both in the spin and orbital sectors ($t$-mixing of all channels) i.e. tunneling   of ($o$) and ($s$) types enriched by  additional  processes  $(\alpha kls)\leftrightarrow (l',-s)$, where $l'\neq l$.
\begin{figure}
\includegraphics[width=0.48\linewidth]{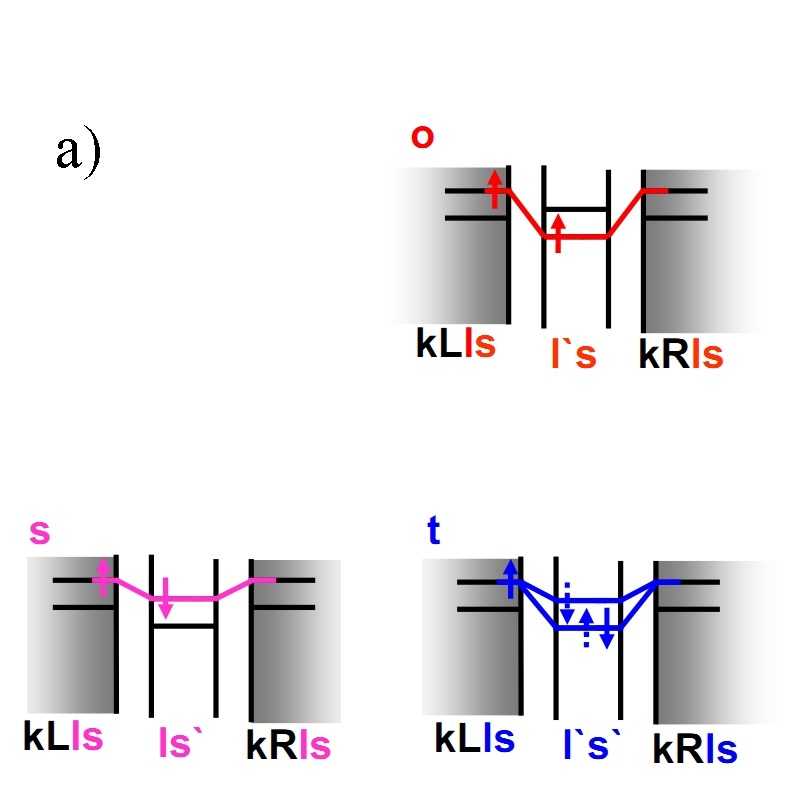}
\includegraphics[width=0.48\linewidth]{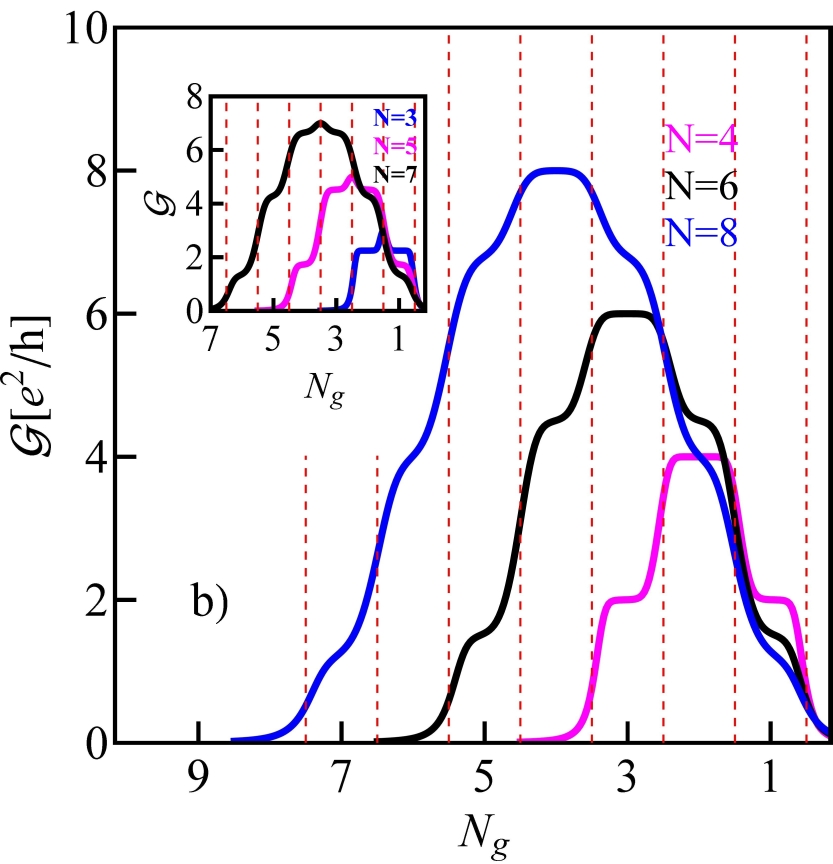}
\caption{\label{fig1} (Color online) Schematic view of orbital ($o$), spin ($s$) and total ($t$) electrode mixing processes. (b) Conductance in the absence of mixing ($nm$, $\nu = 0$) as a function of dimensionless gate voltage $N_{g}$  for even $N$  (main picture) and  odd $N$ ( inset), $\Gamma = 0.01$.}
\end{figure}
To analyze correlation effects, we use finite ${\cal{U}}$ slave boson mean field approach (SBMFA) of Kotliar and Ruckenstein (for the details of the method see e.g.\cite{Krychowski}).
In this approach the effect of Coulomb interactions is effectively replaced  by the interaction of quasiparticles (fermions) with auxiliary bosons, which project the state space onto  subspaces of different occupation numbers. In MFA it leads to the picture of noninteracting  quasiparticles in boson fields. The effective resonant line narrowing factors $z_{ls}$ expressed through mean values of boson operators  and correlation induced shifts of the dot energies $\lambda_{ls}$ are found in self consistent SBMFA equations (minimum of the  free energy \cite{Krychowski}).
As results from symmetry $z_{ls}$ and $\lambda_{ls}$ for $nm$, $t$ and $o^{even}$ cases, are equal for all ($ls$), $z\equiv z_{ls}$, $\lambda\equiv \lambda_{ls}$ (the introduced top notes inform whether an even or odd number of channels are involved in mixing).
For the cases $o^{odd}$ and $s^{odd}$ there are two values for $z$ and $\lambda$: $z_{+}\equiv z_{l+}, z_{-}\equiv z_{l-}$ ($\lambda_{+}\equiv \lambda_{l+}, \lambda_{-}\equiv \lambda_{l-}$) ($o^{odd}$) and for $s^{odd}$: $z_{1}=z_{ls}$ (where $l=1,..(N-1)/2$) and $z_{2}=z_{(N+1)/2+}$.
Analogously, there are also two values of $\lambda$: $\lambda_{1(2)}$. Mixing of electrode channels ${\cal{H}}^{mix}_{d-c}$ results in   the dot states ${ls}$ being mixed. We will number the new basis of independent states on the dot (bonding, antibonding) with the index $m$, $m = 1,...N$.
\begin{figure}
\includegraphics[width=0.48\linewidth]{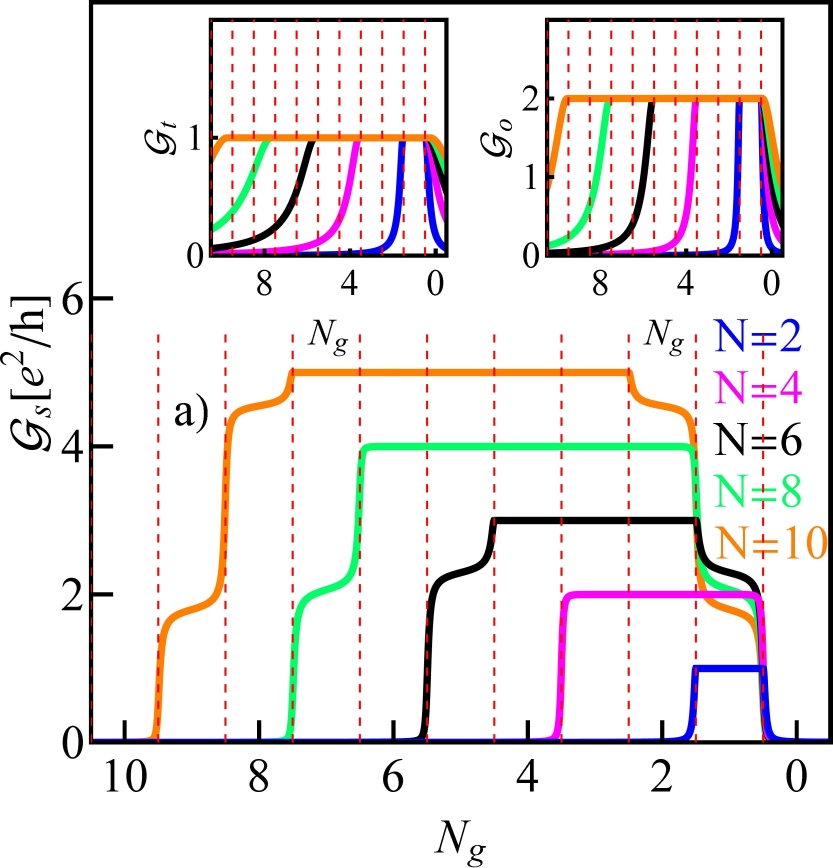}
\includegraphics[width=0.48\linewidth]{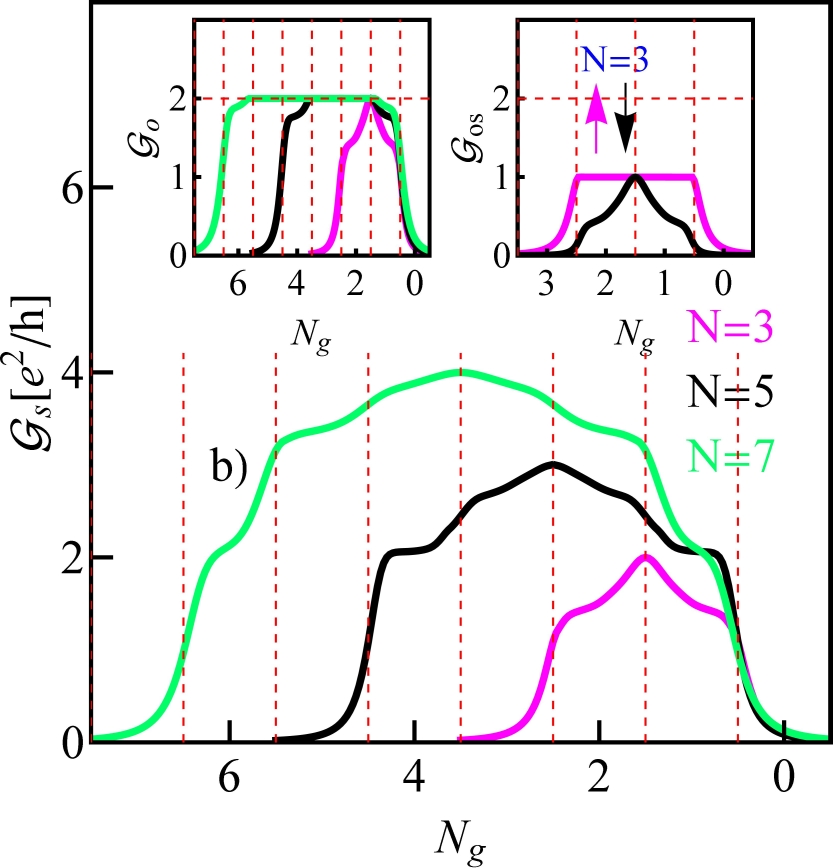}
\caption{\label{fig2} (Color online) The total conductances for $\nu = 1$ versus gate voltage:  a) for the spin mixing ($s^{even}$ for even $N$), left and right insets present the similar conductance curves for ($t$) and ($o^{even}$) respectively. b) Total conductances for odd number of dot states ($s^{odd}$ - main picture, $o^{odd}$ – left inset). The right inset shows spin up and spin down contributions to the conductance for $N = 3$.}
\end{figure}
In this basis conduction (${\cal{G}}$) and thermoelectric power (${\cal{S}}$) can be expressed by transport coefficients as follows:
 ${\cal{G}}=\sum_{m}{\cal{G}}_{m}=\sum_{m}(e^{2}/h){\cal{L}}_{m0}/T$, ${\cal{S}}=\sum_{m}{\cal{S}}_{m}=\sum_{m}(-k_{B}/e){\cal{L}}_{m1}/(T\sum_{m'}{\cal{L}}_{m'0})$, where ${\cal{L}}_{mn=0,1}=\sum_{\alpha}\int^{+\infty}_{-\infty} (E-\mu_{\alpha})^{n}f_{\alpha}(E){\cal{T}}_{m}(E)dE$. $f_{\alpha}(E)$ are the Fermi distribution functions of electrodes and $\mu_{\alpha}=\pm V_{sd}/2$. The $m$-th channel contribution  to the transmission reads   ${\cal{T}}_{m}(E)=(\Lambda_{m}\Delta_{m})/((E-E_{m})^{2}+\Delta_{m}^{2})$, where $\Delta_{m}$ and $E_{m}=E_{d}+\lambda_{m}$ are width and position of many-body resonance. The explicit expressions for $\Lambda_{m}$ and $\Delta_{m}$ are given in Sec. 3.

\section{Results}
\begin{figure}[b!]
\includegraphics[width=0.48\linewidth]{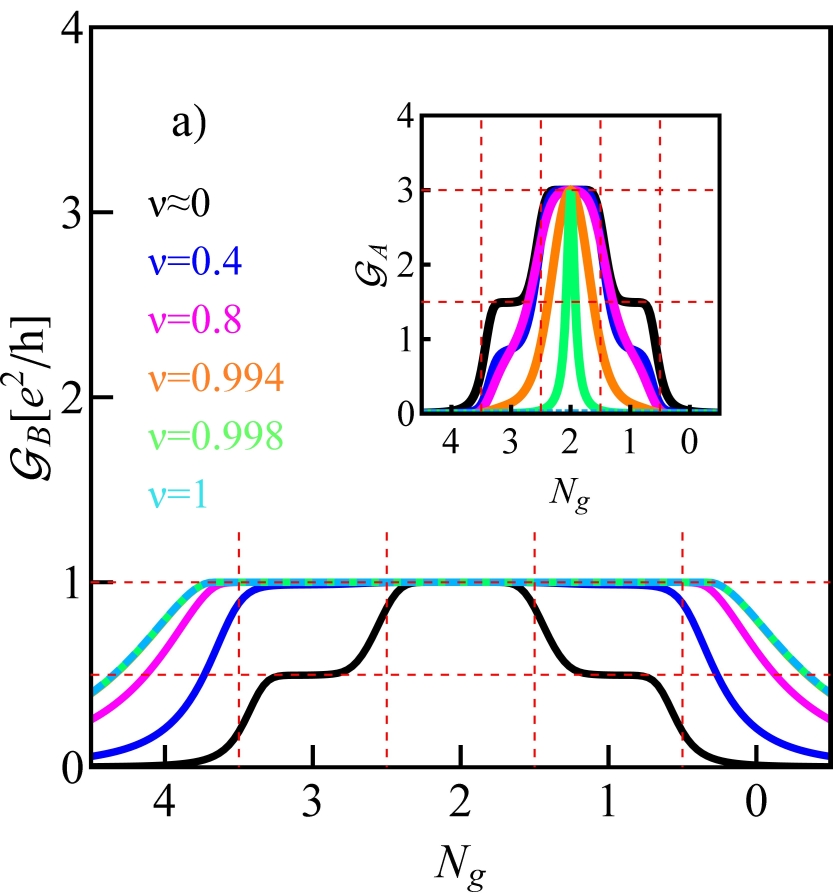}
\includegraphics[width=0.48\linewidth]{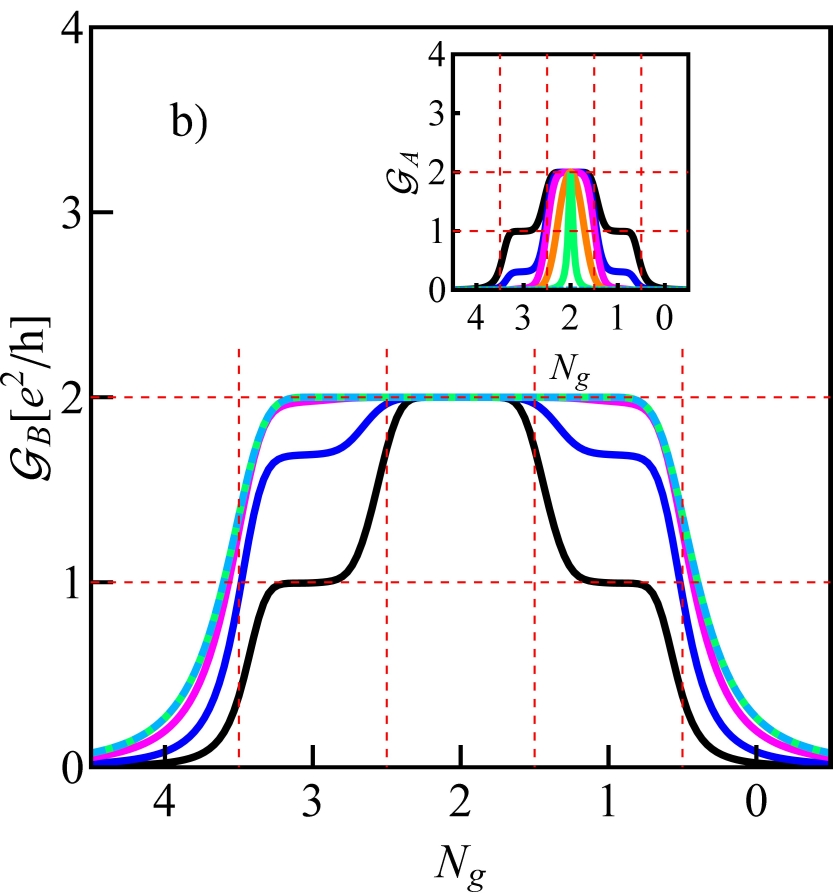}\\
\includegraphics[width=0.48\linewidth]{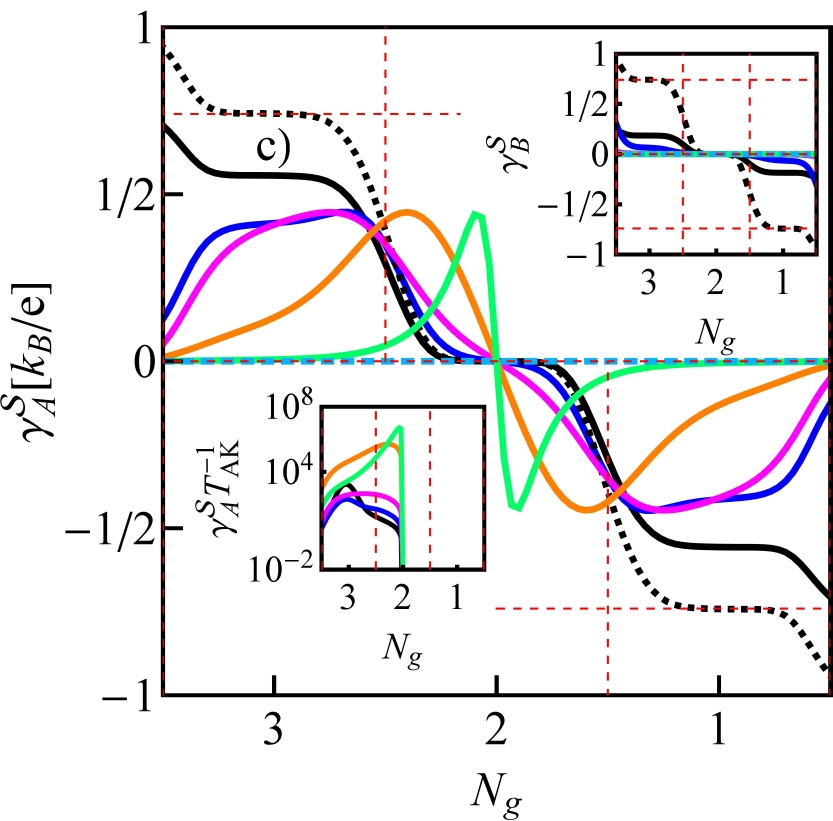}
\includegraphics[width=0.48\linewidth]{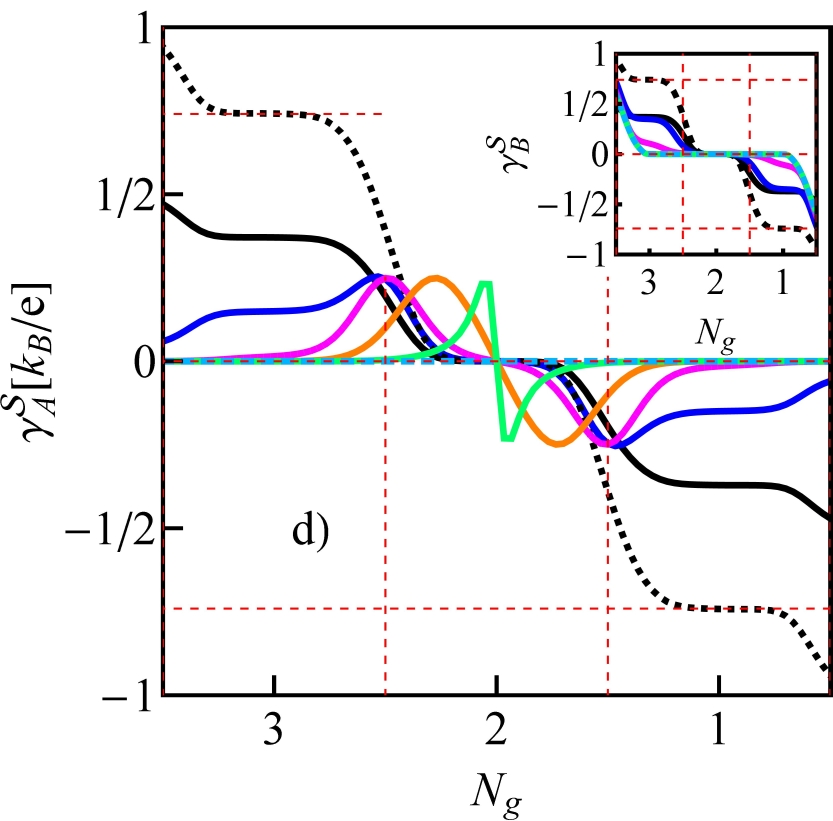}\\
\includegraphics[width=0.48\linewidth]{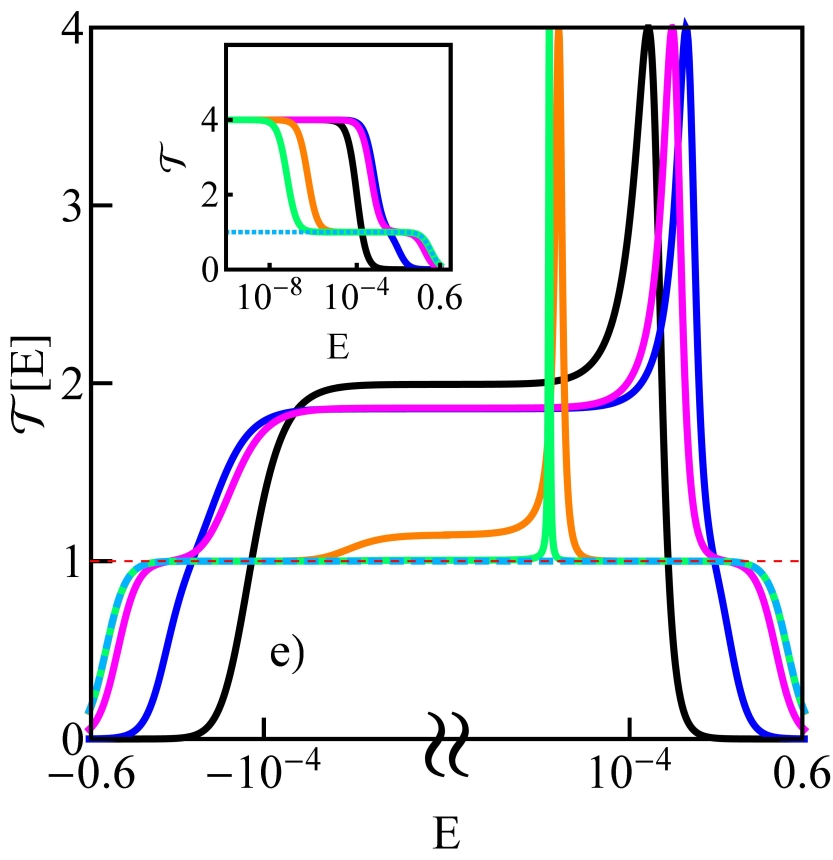}
\includegraphics[width=0.48\linewidth]{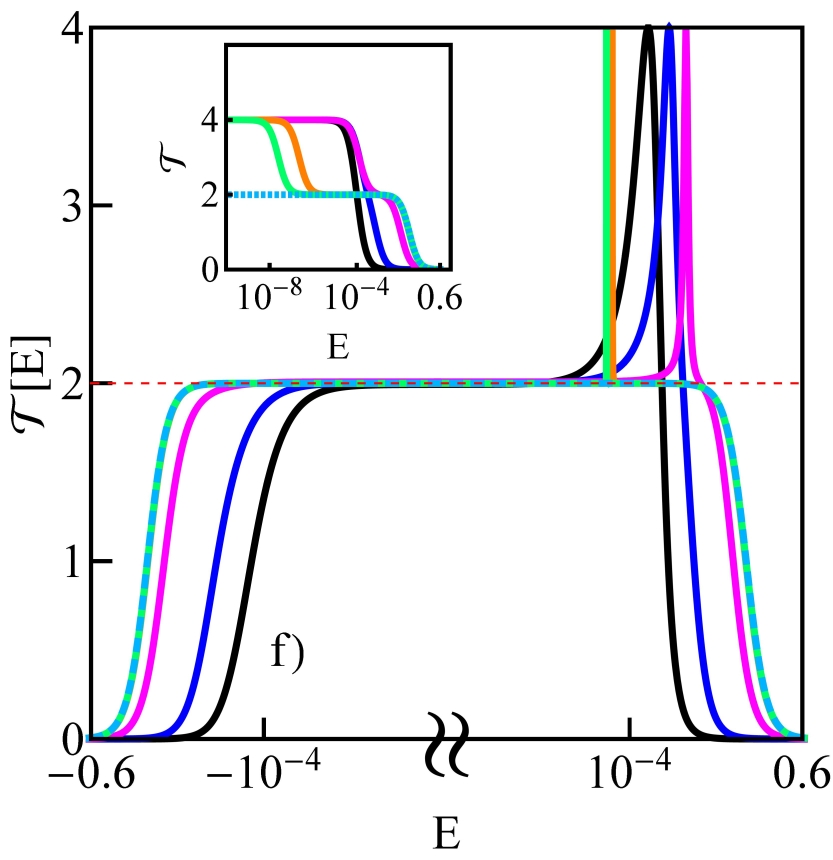}\\
\caption{\label{fig3} (Color online) Conductances for different mixing parameters $\nu$, $N = 4$. a) Bonding state contribution to the conductance (main picture) vs. $N_{g}$ for ($t$) type mixing for different  parameters $\nu$ and corresponding antibonding  contributions (inset).  b) Conductance from B states and in the inset from A states for ($s^{even}$) or ($o^{even}$) cases. Figs. c,d present linear thermoelectric coefficients corresponding to antibonding  and in the upper insets for bonding states for the cases of total – (c) and spin mixing -(d). The left lower inset of (c) shows the example of  giant thermopower (the black dotted curve is the reference line for SU(4) symmetry). e,f) Corresponding total transmission lines for case $t$ and  $s^{even}$ correspondingly for  $N_{g}=1$ and in the insets for $N_{g} = 2$, $\Gamma = 0.03$.}
\end{figure}
One can decompose the total linear  conductance  into separate contributions from bonding (B), antibonding (A) and unmixed states (C).  State (C)  indexed by quantum numbers of isolated dot  ($ls$) occurs only for odd values of $N$. In the following  we will choose as (C) state ($(N+1)/2$,$+$). In the limit of  $V_{sd}\rightarrow0$, $T\rightarrow0$ conductance reads:
\begin{eqnarray}
&&{\cal{G}}_{m}=\frac{e^{2}}{h}\lim\limits_{T,V_{sd}\to 0}\frac{\sum_{\alpha}\frac{\Lambda_{m}}{4\pi}
Re\left[\Psi_{1}\left(\frac{1}{2}+\frac{E_{m}-\mu_{\alpha}+i\Delta_{m}}{2\pi iT}\right )\right]}{T}\nonumber\\&&\approx\frac{e^{2}}{h}\frac{\Lambda_{m}\Delta_{m}}{T^{2}_{mK}}
\end{eqnarray}
where $\Psi_{1}$ is hypergeometric trigamma function.  $ T^{2}_{mK}$ denotes  characteristic resonance temperature calculated in SBMFA, which can be written as $T^{2}_{mK}=E^{2}_{m}+\Delta^{2}_{m}$ \cite{Coleman}. Amplitudes $\Lambda_{m}$ and resonance widths $\Delta_{m}$ depend on the type of mixing. In the case ($t$) there occur ($N-1$) dot antibonding states (A) and one bonding state (B) and then $\Lambda_{A}=(N-1)\Delta_{A}=(N-1)(\nu-1)^{2}\Gamma z^{2}$ and $\Lambda_{B}=\Delta_{B}=((N-1)\nu+1)^{2}\Gamma z^{2}$.  $\Gamma=(2\pi{\cal{V}}^{2})/2D$, where $D$ is the bandwidth. For orbital mixing ($o$) the  situations with an even  and an odd number of dot states should be distinguished.
For $o^{even}$ the amplitudes are given by $\Lambda_{A}=2(N/2-1)\Delta_{A}=2(N/2-1)(\nu-1)^{2}\Gamma z^{2}$ ($\Lambda_{B}=2\Delta_{B}=2(\nu+1)^{2}\Gamma z^{2}$) and for case $o^{odd}$: $\Lambda_{A_{\pm}}=(N_{\pm}-1)\Delta_{A_{\pm}}=(N_{\pm}-1)(\nu-1)^{2}\Gamma z^{2}_{\pm}$ and $\Lambda_{B_{\pm}}=\Delta_{B_{\pm}}=((N_{\pm}-1)\nu+1)^{2}\Gamma z^{2}_{\pm}$.
For the spin mixing, one also has to distinguish between the case of even or odd $N$. For $s^{even}$ case $\Lambda_{A}=(N/2)\Delta_{A}=(N/2)(\nu-1)^{2}\Gamma z^{2}$ and $\Lambda_{B}=(N/2)\Delta_{B}=(N/2)(\nu+1)^{2}\Gamma z^{2}$. For $s^{odd}$: $\Lambda_{A}=((N_{1}-1)/2)\Delta_{A}=((N_{1}-1)/2)(\nu-1)^{2}\Gamma z^{2}_{1}$, $\Lambda_{B}=((N_{1}-1)/2)\Delta_{B}=((N_{1}-1)/2)(\nu+1)^{2}\Gamma z^{2}_{1}$ and the amplitude of the unmixed state $\Lambda_{C}=\Gamma z^{2}_{2}$.

In the limit $V_{sd}\rightarrow0$ and $T\rightarrow0$ the linear temperature coefficient of thermopower defined by $\gamma^{{\cal{S}}}_{m}=\frac{{\cal{S}}_{m}T_{mK}}{2\pi T}$ takes  the value:
%\lipsum[1]
\begin{widetext}
\begin{equation}
%\[
\gamma^{{\cal{S}}}_{m}=\lim\limits_{T,V_{sd}\to0}\frac{-k_{B}T_{mK}Im\left[\sum_{\alpha}\frac{\Lambda_{m}\left(E_{m}-\mu_{\alpha}+i\Delta_{m}\right)}{4\pi i}\Psi_{1}\left(\frac{1}{2}+\frac{E_{m}-\mu_{\alpha}+i\Delta_{m}}{2\pi iT}\right )\right]}{e2\pi T^{2}\sum_{\alpha m'}\frac{\Lambda_{m'}}{4\pi}Re\left[\Psi_{1}\left(\frac{1}{2}+\frac{E_{m'}-\mu_{\alpha}+i\Delta_{m'}}{2\pi iT}\right )\right]}\approx\frac{-k_{B}\pi E_{m}\Delta_{m}\Lambda_{m}\prod_{n\neq m}^{N-1}T^{2}_{nK}}{3eT_{mK}\sum_{m}^{N}\Delta_{m}\Lambda_{m}\prod_{n\neq m}^{N-1}T^{2}_{nK}}
%\]
\end{equation}
\end{widetext}
%\lipsum[1]
where ${\cal{S}}_{m}$ denotes the electron contribution to the  thermopower from state $m$. For $\nu=0$ the above formula takes the form $\gamma^{{\cal{S}}}=\gamma^{{\cal{S}}}_{A}+\gamma^{{\cal{S}}}_{B}=-(k_{B}/e)(\pi/3)E_{N}/T_{NK}$ where $E_{N}$ and $T_{NK}$ are position of the resonance  and Kondo temperature  for SU(N) respectively. We pay special attention to $\gamma^{{\cal{S}}}$ coefficient, because this quantity similar to the  conductance has distinct plateaus in the range of strong Kondo correlations (the examples for SU(4) are  shown on Figs. 3c,d). From the conductivity measurements one  gets information about  Kondo temperature, while  $\gamma^{{\cal{S}}}$ coefficient supplements information about resonance by  specifying energy location of the resonance. Using SBMFA expressions on $E_{N}$ and $T_{NK}$: $E_{N}=\Delta_{N}cot(\pi n)$ and $T_{NK}=\Delta_{N}/|sin(\pi n)|$ \cite{Coleman}, $\gamma^{{\cal{S}}}$ can also be written in the form $\gamma^{{\cal{S}}}=-(k_{B}/e)(\pi/3)cos(\pi n)$, where $n$ denotes the occupation number. For the special case of SU(4) symmetry this result  has already been derived earlier in \cite{Bas}. For $\nu=1$ $\gamma^{{\cal{S}}}=\gamma^{{\cal{S}}}_{B}=-(k_{B}/e)(\pi/3)E_{B}/T_{BK}$ and at the point of electron-hole symmetry (e-h) ($E_{d}=-(3/2){\cal{U}}$) $\gamma^{{\cal{S}}}=0$.
\begin{figure}
\includegraphics[width=0.48\linewidth]{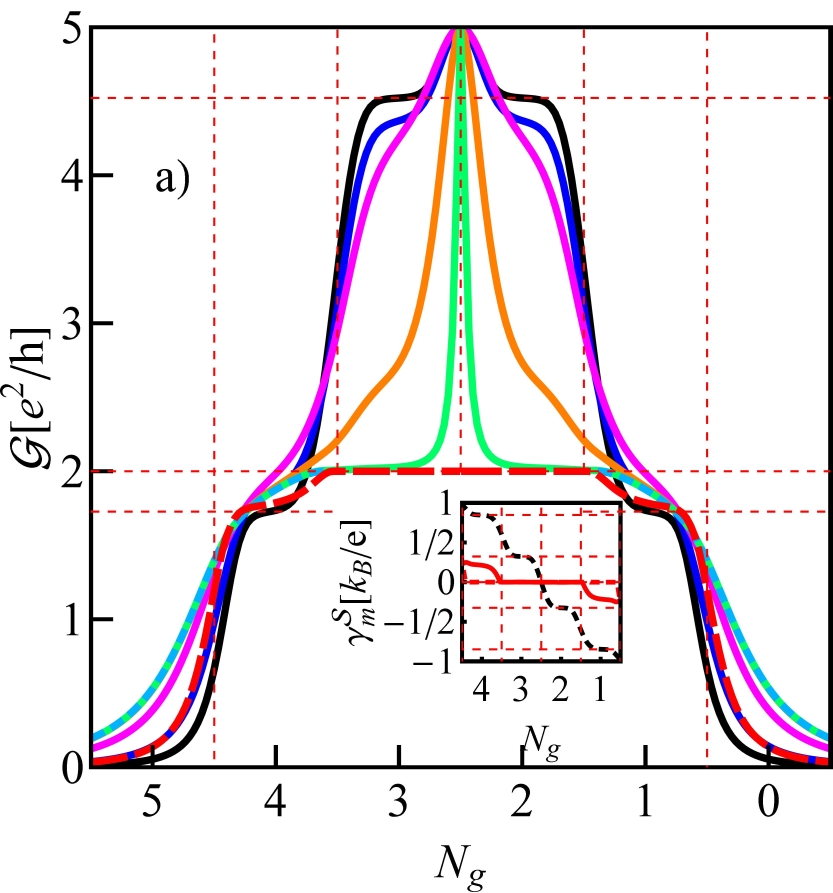}
\includegraphics[width=0.48\linewidth]{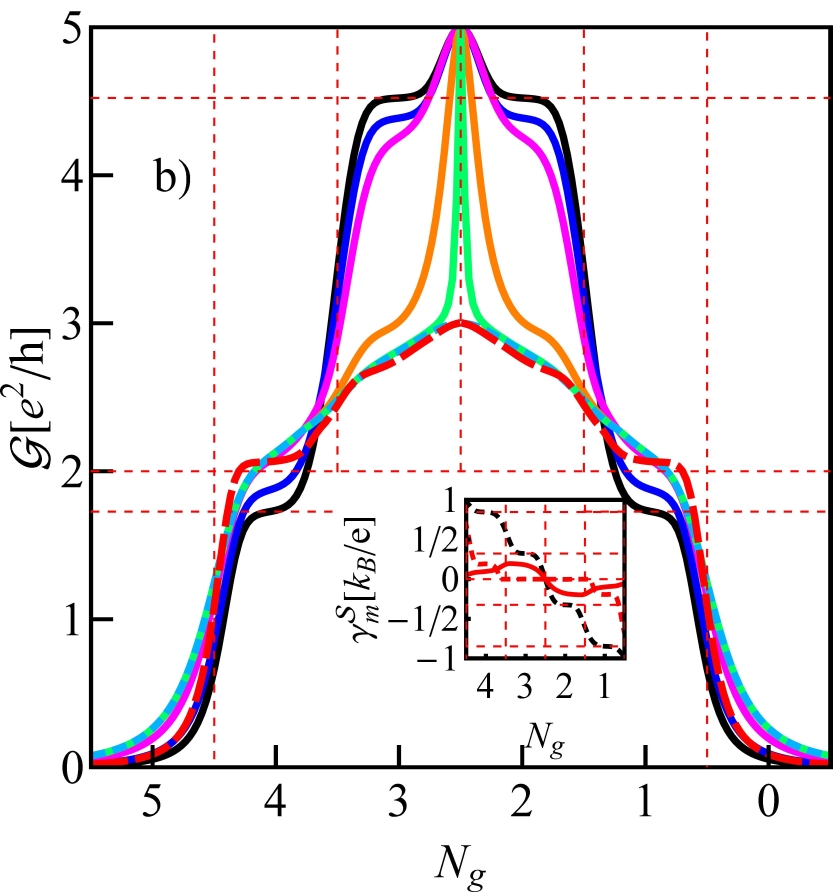}\\
\caption{\label{fig4} (Color online) Conductances for the same choice of mixing parameters as in Fig. 3, here for $N = 5$: a) orbital mixing ($o^{odd}$), b) spin mixing ($s^{odd}$), $\Gamma =0.03$. The red solid lines are drawn as the references (case $\nu =1$, $\Gamma =0.02$). Insets present $\gamma^{{\cal{S}}}_{m}$ for orbital and spin mixing respectively ($\nu = 0$ – black line,  $\nu=1$ – red lines: for left inset $m = B2$ (solid), $m = B1$ (dashed), and for right $m = C$ – solid, $m = B$ – dashed.}
\end{figure}

The numerical results discussed below are presented with the use of energy unit $D/50=1$ and we take ${\cal{U}}=3$. It is convenient to work with a dimensionless gate voltage defined by $N_{g}=(1/2)(1-2E_{d}/{\cal{U}})$. This quantity approximately describes occupation regions.

Fig. 1b shows examples of gate voltage dependencies of  conductance of fully symmetric systems SU(N) ($\nu=0$) for even number of dot states $N=4,6,8$ and in the inset  for odd values $N=3,5,7$.  Clearly seen  successive plateaus are the manifestations of SU(N) Kondo effects in different occupation regions. For the currently analyzed case of no mixing, the values of conductance are dictated by Friedel sum rule ${\cal{G}}=Nsin^{2}\left( \frac{\pi n}{N}\right)$. For even $N$, Kondo resonance at the e-h symmetry point locates at the Fermi level and the corresponding Kondo temperature is the lowest of the entire occupation  range and conductance is the highest and takes the value $N(e^{2}/h)$. Maximum of conductance visible for odd $N$ places also on the e-h symmetry point and  has a value $N(e^{2}/h)$, but in this case it locates at the border of the Coulomb blockade.

Fig. 2 presents  conductances for the opposite  case of full mixing ($\nu=1$) of different types. Fig. 2a concerns $s^{even}$ mixing. As a result of spin mixing one bonding state (B) and one antibonding state (B) are formed in each of the $N/2$ orbital sectors. For $\nu=1$ only the first type of states contributes to the conductance and therefore it takes the value $e^{2}/h$, which is the value for SU(N/2). For the special case of $N = 4$ this effect  has already been described earlier in \cite{Martins} as a symmetry reduction SU(4)$\rightarrow$ 2LSU(2)(two-level SU(2)). The effect discussed  here is a generalization: SU(N)$\rightarrow$ (N/2)LSU(N/2). Conductances for  the other two types of mixing are shown in the insets (left inset - ($t$) mixing, right - ($o$)). In the case ($t$), ($N-1$)  A states and one B state are formed and only the latter contributes to the conductance for $\nu=1$. This wide resonance  centered at the Fermi level is a result of joint action of  cotunneling and interference processes and gives unitary contribution, equal independent of $N$ value (${\cal{G}}=2(e^{2}/h)$).  The plateau occurring in a wide range of gate potential indicates the correlative nature of this resonance. For the case $o^{even}$ in each of spin sectors  $(N/2-1)$ A states are formed and one state B. Unitary contribution to the conductance give two degenerate B states with opposite spins  ${\cal{G}}=2(e^{2}/h)$. Here we encounter also 2LSU(2) Kondo effect.

Fig. 2b  illustrates the case of odd numbers $N$ ($\nu=1$).  For $s^{odd}$ there are $(N-1)/2$ A states and the same number of B states and additionally appears unmixed state (C). This time not only B states, but also C state  contribute to the conductance. As an example we show in the right inset the partial conductances, where it is seen that  B state contribution exhibits wide plateau, whereas gate dependence of C maps the shape of the output symmetry SU(3), of course with different limits of conductance.  Due to the lack of mixing, state C retains the memory of the original symmetry. For the orbital mixing case ($o^{odd}$ - left inset), the two spin sectors are not equinumerous, $(N+1)/2$ states with spin up and  $(N-1)/2$ states with spin down. As a consequence, two types of bonding states and two types of antibonding states appear. In the areas $N_{g}\approx1$ and $N_{g}\approx(N-1)$  two bonding states are nondegenerate. Lower on the energy scale gives  a unitary contribution to the conductance and higher gives less than unity. With the increase of $N$, however,  the energy difference between both states decreases and then also the latter  contribution reaches the unitary limit. In other occupation areas two bonding states are degenerate and conductance is unitary. It is worth to emphasize  that while in the case of 2LSU(2) resonance, partial transmissions  from both orbitals are identical in the entire energy range, in the case now under discussion, the lines from both degenerate orbitals differ, they are centered on the same energy value, but their  widths are different.

Figures 3 and 4 present evolution of conductance and linear temperature coefficient of thermopower with the increase of mixing. Fig. 3 shows examples for even $N$($N = 4$) and Fig. 4 for odd ($N =5$). These cases distinctly  differ, which was already evident in the limit $\nu\rightarrow1$. As we have mentioned for even $N$ and  $\nu = 1$ only one bonding state is involved in the transport, whereas for odd $N$ two bonding states or one bonding and one unmixed state. Fig. 3a shows partial conductance related to the bonding states (${\cal{G}}_{B}$) and in the inset contribution from antibonding (${\cal{G}}_{A}$). It is seen that with the increase of $\nu$ the gate dependence of ${\cal{G}}_{B}$ loses the shape of the output symmetry and partial conductance  evolves to  the unitary limit associated with the domination of the  state B.  The gate dependence of antibonding contribution (inset of Fig. 3a) for $\nu = 0$ starts from a shape typical for the output symmetry and gradually quenches with the increase of $\nu$, most effectively  further from the e-h symmetry  point. Removal of degeneracy caused by  mixing is weakest close to e-h symmetry  point and in this region ${\cal{G}}_{A}$ maintains the value  corresponding to the  fully symmetrical system even for $\nu$ close to 1.  Fig. 3b illustrates similar process, this time with mixing in the spin or orbital sector,  which for $N = 4$ is equivalent,  because spin and orbital pseudospin have the same dimension in this case. One can see, that the bonding contribution approaches value $2(e^{2}/h)$ due to participation of two B states and states A gradually contribute less and less to the conductance and disconnect for $\nu = 1$ (2LSU(2)\cite{Martins}). Fig. 3c,d supplement information on transport properties by presentation of thermopower coefficients: bonding contribution $\gamma^{\cal{S}}_{B}$ and antibonding $\gamma^{\cal{S}}_{A}$ in the insets. Due to the centering of the bonding resonance on $E_{F}$ with the increase of  $\nu$, also in $N_{g}\approx2$ areas   $\gamma^{{\cal{S}}}_{B}\rightarrow0$ (see Fig. 3f,g  illustrating evolution of  transmission  with the change of  $\nu$). $\gamma^{\cal{S}}_{A}$ gradually disappears in the $N_{g}\approx1,3$ regions , because A resonance itself  disappears. In the region $N_{g}\approx2$ the maximum of  $\gamma^{\cal{S}}_{A}$ is clearly marked $\gamma^{{\cal{S}}}_{A}=1/\left(1+\frac{\Lambda_{B}\Delta_{B}T^{2}_{AK}}{\Lambda_{A}\Delta_{A}T^{2}_{BK}}\right)^{2}$, which is associated with narrowing of the resonance peak, for $\nu = 1$ A state sharply disconnects ($\gamma^{\cal{S}}_{A}=0$ )(insets of Figs. 3e,f). It is worth mentioning that thermopower  reaches  gigantic values in the point, where  $\gamma^{\cal{S}}_{A}$ has its maximum (left down inset on Fig. 3c). This is a consequence of the fact that for this energy the resonance peak is extremely narrow and its distance from $E_{F}$ is smaller than the width.

Figs. 4a,b illustrate evolution of   conductance for $N = 5$ with the increase of mixing. For orbital mixing (Fig. 4a) conductance in the regions $N_{g}\approx2,3$ decreases from the value characteristic for SU(5) symmetry $5(5/8+\sqrt{5}/8)(e^{2}/h)$ to  value $2(e^{2}/h)$, which is dictated by the participation of two degenerate bonding states.  In the regions  $N_{g}\approx1,4$ conductance increases from $5(5/8-\sqrt{5}/8)(e^{2}/h)$. This is caused by lifting of degeneracy of the bonding states. The energetically lower bonding state  (B1)  moves closer to $E_{F}$ and this  state makes a major contribution to the conductance ($e^{2}/h$). The second bonding state (B2) has a smaller contribution. Due to centering of  B1 resonance line at the Fermi level $\gamma^{{\cal{S}}}_{B1} = 0$, while $\gamma^{{\cal{S}}}_{B2}\neq0$(inset of Fig. 4a). For $N\rightarrow\infty$ and $\nu\rightarrow1$ also B2 resonance line locates  on Fermi level and then total  conductance approaches  unitary limit $2(e^{2}/h)$ and $\gamma^{{\cal{S}}}_{B1}=\gamma^{{\cal{S}}}_{B2}=0$.  For comparison we also present $\gamma^{{\cal{S}}}$ for the  fully symmetric case  SU(5) and it exhibits  plateuas with the  values characteristic for a given symmetry and occupation regions: $\frac{\mp\pi(\pm1+\sqrt{5})}{12}$ (see Eq. 3).  Case $s^{odd}$ ($N = 5$)  illustrated on  Fig. 4b  differs from the situation presented on Fig. 4a  in the presence of two bonding states and one unmixed state. In the regions $N_{g}\approx2,3$  resonance lines from two bonding states center on the Fermi level for $\nu\rightarrow1$ (${\cal{G}}_{B}=2(e^{2}/h)$, $\gamma^{\cal{S}}_{B}=0$), and the line corresponding to the  unmixed state C centers at $E_{F}$ only at  e-h symmetry point and there it contributes to the conductance (${\cal{G}}_{C}=(e^{2}/h)$ and $\gamma^{\cal{S}}_{C}=0$). In the regions  $N_{g}\approx1, N-1$  both bonding and unmixed lines are shifted from the Fermi level, which is visible in both the conductivity and $\gamma$ coefficients. For $N\rightarrow\infty$ and $\nu = 1$ only bonding states contribute to the conductance in these regions.

Summarizing, in the present paper we have derived the  general expressions for the  linear conductance and linear temperature coefficient of thermopower for strongly correlated multilevel quantum dot  in the case of spin-flip assisted or orbital mixing tunneling and analyzed evolution of transport properties with the degree of mixing. For the fully symmetrical systems SU(N) both conductance and thermopower coefficient  show in strong correlation regimes  plateaus with characteristic values for a given symmetry and occupation number. In the case of full mixing, observed universal values of conductance are dictated by the number of states active in transport and this in turn depends on  the type of mixing and parity of the number of states of the dot. We suggest that close to the  full mixing one can expect a gigantic value of thermopower.

\end{document}